\documentclass[a4paper,aps,prl,showpacs,twocolumn,superscriptaddress]{revtex4-1}   %APS document class style  (PRA,PRL,ETC)
   
\usepackage[utf8]{inputenc}  %Input what you want e.g., é, ł, a, ü
\usepackage[T1]{fontenc}     %Output what you want e.g., é, ł, a, ü
\usepackage[british]{babel}  %Do hyphenation according to british english
\usepackage{lmodern}  %Joe's font
\usepackage[scaled=1.03]{inconsolata} %Monospace font
\usepackage[usenames,dvipsnames]{color} %More color choices
\usepackage[colorlinks,citecolor=blue,linkcolor=magenta,urlcolor=blue]{hyperref}  %Hyperlinks (pink, green, blue)
\usepackage{graphicx} % Package to insert external figures
\usepackage{tikz}
\usepackage[babel]{microtype}  %Improves text justification
\usepackage{amsmath,amssymb,amsthm,bm,mathtools,amsfonts,mathrsfs,bbm,dsfont} %Useful math packages
\usepackage{xspace}  %Useful to add space in macros
\usepackage{multirow}
\usepackage{verbatim}
\usepackage{physics}
\newcommand{\id}{\ensuremath{\openone}}
\usepackage{bbold}

\usepackage{paralist}% In line lists

\newcommand{\<}{\langle}
\renewcommand{\>}{\rangle}
\newcommand{\hil}{\mathcal{H}}
\newcommand{\kil}{\mathcal{K}}
\newcommand{\R}{\mathbb{R}}
\newcommand{\proj}[1]{\left| #1 \middle \rangle \! \middle\langle #1 \right|}
\renewcommand{\tr}[2][\!]{\mathrm{tr}_{#1}\left[ #2 \right]}
\newcommand{\mI}{\mathcal I}% notation for instruments

\newtheoremstyle{mystyle}% name
  {6pt}%Space above
  {6pt}%Space below
  {\normalfont}%Body font
  {0pt}%Indent amount
  {\bf}% Theorem head font
  {.}%Punctuation after theorem head
  { }%Space after theorem head 2
  {}%Theorem head spec (can be left empty, meaning ‘normal’)

\theoremstyle{mystyle}
\newtheorem{theorem}{Theorem}
\newtheorem{lemma}{Lemma}

\newtheorem{corollary}{Corollary}

\begin{document}

\title{All quantum resources provide an advantage in exclusion tasks}
%Characterising the `quantumness' of devices through worst case scenarios}

\author{Roope Uola}
\affiliation{D\'{e}partement de Physique Appliqu\'{e}e, Universit\'{e}  de Gen\`{e}ve, CH-1211 Gen\`{e}ve, Switzerland}

\author{Tom Bullock}
\affiliation{QTF Centre of Excellence, Turku Centre for Quantum Physics, Department of Physics and Astronomy,
University of Turku, FI-20014 Turun yliopisto, Finland
}

\author{Tristan Kraft}
\affiliation{Naturwissenschaftlich-Technische Fakult\"at, Universit\"at Siegen, Walter-Flex-Str. 3, D-57068 Siegen, Germany}

\author{Juha-Pekka Pellonp\"a\"a}
\affiliation{QTF Centre of Excellence, Turku Centre for Quantum Physics, Department of Physics and Astronomy,
University of Turku, FI-20014 Turun yliopisto, Finland
}

\author{Nicolas Brunner}
\affiliation{D\'{e}partement de Physique Appliqu\'{e}e, Universit\'{e}  de Gen\`{e}ve, CH-1211 Gen\`{e}ve, Switzerland}

\begin{abstract}
A key ingredient in quantum resource theories is a notion of measure. Such as a measure should have a number of fundamental properties, and desirably also a clear operational meaning. Here we show that a natural measure known as the convex weight, which quantifies the resource cost of a quantum device, has all the desired properties. In particular, the convex weight of any quantum resource corresponds exactly to the relative advantage it offers in an exclusion task. After presenting the general result, we show how the construction works for state assemblages, sets of measurements and sets of transformations. Moreover, in order to bound the convex weight analytically, we give a complete characterisation of the convex components and corresponding weights of such devices.
    %Recent developments have shown that any quantum resource theory with a convex set of free states allows for the characterisation (quantification) of the resource through its outperformance with respect to the free set in a tailored discrimination task. We show that these resources can equally be characterized well be characterised through their performance in the task of state exclusion. As in the case of outperformance, where the generalised robustness measure forms a quantifier for the resources, in our scenario the corresponding quantifier is the convex weight. The convex weight measures how large a fraction of a given device can be generated by free resources, whereas the generalised robustness measures how much a device can tolerate mixing with abstract operators before losing the resource. As explicit examples of our technique we show how the construction works for state assemblages, measurement assemblages and for sets of transformations. Moreover, in order to bound the convex weight analytically, we give a complete characterisation of the convex components and corresponding weights of such devices.
\end{abstract}

\maketitle

\textit{Introduction.---} Quantum theory allows for concepts that have no analogue in classical physics. Most prominent examples include entangled states, incompatible measurements, and quantum memories. An important question is to characterize these genuinely quantum resources, in particular to quantify their non-classicality. A natural approach to this problem is to view these genuine quantum properties as resource for some task, and ask to what extent a given quantum device deviates from the classical scenario. Recently, a general framework of quantum resource theories has been developed to address these questions (see Ref. \cite{Gour2019} for a recent review). These ideas have already been formally applied to a broad range of quantum properties, such as entanglement \cite{1509.07458}, joint measurability \cite{1502.03010,1908.11274}, steering \cite{Piani2015,PhysRevX.5.041008}, thermal operations \cite{1805.09564}, asymmetry \cite{1209.0921} and coherence \cite{PhysRevLett.116.120404}. 
	
In general, a resource theory is defined via a set of free resources (for instance associated to classical resources), and a set of free operations. Applying a free operation to a free resource should always give back a free resource, and more generally free operations cannot boost the available resource. Hence, classical pre- and post-processings are usually part of free operations, which implies that the set of free resources must be convex. This motivates the use of convexity-based measures in order to quantify quantum resources, i.e. to measure their non-classicality. 

Recently, a large body of work has been devoted to one of these measures, namely the generalised robustness \cite{Piani2015,uola2015one,PhysRevLett.116.150502,1906.00448,Ryuji2019,TaRe2019,Michal2019,CaHeTo2019,SkSuCa2019,uola2019a,uola2019b}. The latter quantifies the resource of a given device, by asking by how much it can be mixed with another (arbitrary) device before the resource is lost (i.e. the mixture belongs to the free set). Loosely speaking, this captures the distance between a given device and the set of free devices. Since its introduction, the generalized robustness has been found to possess three very attractive and fundamental properties: $(i)$ faithfulness, i.e. it is zero if and only if a device is free, $(ii)$ convexity, $(iii)$ monotonic under free operations, $(iv)$ it quantifies the outperformance of a quantum device with respect to all classical ones in an explicit task, namely a discrimination game, (v) it can be calculated efficiently when the free set can be expressed through semi-definite constraints (thereby forming a certificate).

In this manuscript, we prove that another, also well-motivated, quantifier has the five fundamental properties mentioned above. This quantifier is known as the convex weight. It has a natural interpretation in the context of resource theories.   
Namely, it characterizes how large fraction of a given resource device can be generated with free (or classical) resources. 
In this sense, the convex weight provides a direct quantifier of the resource cost, and is thus complementary to the generalized robustness. Consider for instance a resource that is extremal, but very close to the free set. While the generalized robustness is very small for this resource, the convex weight will nevertheless be equal to zero. 
	
In order to prove property (iv), we construct explicitly a task for which the convex weight quantifies exactly the relative advantage provided by the resource over any free device. This task corresponds to an exclusion (or anti-distinguishability) task. That is, given a randomly chosen element $x_k$ from a known list of elements $\{x_i\}$, one should provide as answer any $x_i \neq x_k$. 
%(?? This exclusion task can be seen as the converse to a discrimination game, the latter being naturally associated the generalized robustness.) 
After discussing the general framework, we discuss the cases where the quantum devices corresponds to sets states, sets of measurements and quantum channels. For instance, any set of incompatible quantum measurements provides an advantage in a task of state exclusion, and the convex weight represents the relative advantage over any set of compatible (i.e. jointly measurable) measurements. Finally, we show that the convex weight can be easily bounded (and in simple cases even decided) by fully analytical methods, a fact that we illustrate by characterising all devices and corresponding weights that can appear in a convex decomposition of a given device.

%The proposed quantifier is applicable to various free sets of devices. For example, one can choose unsteerable states, compatible measurements, entanglement breaking channels and finite round protocols based on local operations and classical communication (LOCC). Since the quantifier requires only control of the input state and the output measurements, it is experimentally suitable. Examples of possible tests that have been performed in this direction can be found in \cite{Sun18,Zheng2018}.

\textit{Convex weight of quantum devices.---}
We concentrate on three categories of quantum devices: quantum states, measurements and transformations. 
We may also extend the notion to include sets of such devices, e.g., a collection of states or measurements.
Formally, states correspond to positive unit-trace operators denoted by $\varrho$, measurements to collections of positive operators denoted by $\{M_j\}$ with the normalisation $\sum_j M_j=\id$, and transformations are given as completely positive trace non-increasing maps denoted by $\mI$ (or $\Lambda$ if trace preserving). For a convex and compact subset $F$ of (a given class of) quantum devices one defines the convex weight $\mathcal{W}_F(D)$ of a device $D$ as the maximum relative number of times a device from the set $F$ can be used to produce $D$. Formally,
\begin{equation}
\label{WeightDev}
\begin{split}
    \mathcal{W}_F(D) = &\min \lambda\\
    &\mathrm{s.t.:}\  D= (1-\lambda) D_F + \lambda\tilde D,
\end{split}
\end{equation}
where the optimisation runs over devices $D_F$ in the subset $F$ and general devices $\tilde D$ outside of $F$.

The weight has the following appealing properties of a resource quantifier: $(i)$ faithfulness, i.e. it is zero if and only if a device is free, $(ii)$ convexity, $(iii)$ monotonic under free operations, $(iv)$ task-oriented interpretation, $(v)$ simple to bound analytically. The properties $(i)$ and $(iii)$ follow directly from the definition, the properties $(ii)$ and $(v)$ are proven in the Appendix and the property $(iv)$ is the main message of this manuscript. We note that as a by-product of proving property $(v)$ we provide a complete characterisation of the convex components of a given device.

\textit{Exclusion input-output games.---}
An input-output game $\mathcal{G}$ is defined as a triplet $\mathcal{G}=(\mathcal{E},\mathcal{M},\Omega)$, where $\mathcal{E}=\{p(i)\varrho_i\}$ is a state ensemble, $\mathcal{M}=\{M_j\}$ forms a POVM, and $\Omega=\{\omega_{ij}\}$ is a real-valued reward function. The task is to find a transformation $\mI$ that minimises the payoff defined as
\begin{align}
    P(\mathcal I,\mathcal{G}):=\sum_{ij}\omega_{ij} \ p(i)\tr{\mI(\varrho_i)M_j}.
\end{align}
Note that in the case $\mI=id$ and $\omega_{ij}=\delta_{ij}$ the payoff corresponds to the exclusion probability in a minimum error state discrimination task. Note also that in contrast to input-output games, where the task is to maximise the payoff, in exclusion input-output games we are interested in minimisation. One could argue that there is not much difference between input-output games and their exclusion variants, as one can flip the signs in the reward functions and look at the absolute value of the payoff. The difference between the games becomes evident, however, when looking at canonical input-output games (see below) that remove all covariance between the payoff and the reward function. As it turns out, such elimination of covariance is necessray for the connection between resource measures and quantum games \cite{1907.02521,uola2019b}. This duality between the games also highlights the duality between the concepts of generalised robustness and the convex weight.

In order to define input-output games for sets of devices, we define the games for each device separately and as payoff we take the sum of the individual payoffs. Formally, in the definition of a game we replace the state ensemble $\mathcal{E}=\{p(i)\varrho_i\}$ with a state assemblage $\mathcal{A}=\{p(i,x)\varrho_{i|x}\}$, the single POVM $\mathcal{M}=\{M_j\}$ with a measurement assemblage $\mathcal{M}_A=\{M_{j|x}\}$, and the reward function $\Omega=\{\omega_{ij}\}$ with a fine-tuned reward function $\Omega_f=\{\omega_{ijx}\}$. We refer to the triplet $(\mathcal{A},\mathcal{M}_A,\Omega_f)$ with the same symbol $\mathcal{G}$ as used above when there is no risk of confusion. Now the payoff for a set of transformations $\{\mI_x\}$ reads
\begin{equation}
    P(\{\mI_x\},\mathcal{G}):=\sum_{ijx}\omega_{ijx} \ p(i,x)\tr{\mI_x(\varrho_{i|x})M_{j|x}}.
\end{equation}
Again, the case $\{\mI_x\}=\{id\}$ and $\omega_{ijx}=\delta_{ij}$ corresponds to a minimum error discrimination task.

Input-output games that do not relate to minimum error discrimination have some redundancy: a game can be transformed into another one by scaling the reward function or by adding a constant to it. This results in a scaled or shifted payoff. In order to treat such games on an equal footing, we eliminate the scaling and shifting covariance by defining canonical versions of the games. A canonical version of a game is obtained by first shifting the lowest payoff to zero when optimised over (sets of) transformations and then scaling the highest payoff to one.

\textit{Main idea.---}
The convex weight $\mathcal{W}_F(D)$ of a device $D$ with respect to a free set $F$ is defined in Eq.~(\ref{WeightDev}). Solving this equation for $\tilde D$ and defining $\hat D_F:=(1-\lambda) D_F$ results in
\begin{eqnarray}
\label{WeightDev2}
\mathcal{W}_F(D) = \min\, && \lambda \\
\text{s.t.: }&& \frac{1}{\lambda}(D-\hat D_F)\in\text{Dev},\quad \hat D_F\in C_F, \notag
\end{eqnarray}
where $\text{Dev}$ is the set of all devices and $C_F=\{\alpha D_F|\alpha\geq 0, D_F\in F\}$ is a cone based on the subset $F$.

Of the two optimisation constraints the conic one is linear and it reads the same for all three categories of devices (or sets thereof). The other constraint, however, is non-linear and as such we check it in more detail. For states and measurements the constraint reduces to positive semi-definiteness of $D-\hat D_F$. This is due to the fact that the normalisation, i.e. having unit trace or a sum equal to identity, of $\frac{1}{\lambda}(D-\hat D_F)$ is automatic. A transformation can be seen as an element of a quantum instrument, i.e. a collection of completely positive trace non-increasing maps summing to a completely positive trace preserving map. As such, we consider instruments in place of transformations from here on. This way the non-linear optimisation constraint becomes linear. Namely, as the normalisation is now automatic, the operator $\frac{1}{\lambda}(D-\hat D_F)$ being an instrument corresponds to positive semidefiniteness of the Choi picture of $D-\hat D_F$.

With the above modification we have brought the problem of calculating the weight of a quantum device $D$ with respect to the set $F$ into a linear problem with conic constraints. Such optimisation problems are called conic programs. Below we spell out these programs explicitly in their dual form for states, measurements and channels as a special case of instruments. The treatment of general instruments is presented in the Appendix. The connection between convex weight and the performance in input-output games follows directly from the dual.

For a state assemblage $\mathcal{A}=\{p(i,x)\varrho_{i|x}\}$ the primal of the cone program \eqref{WeightDev2} reads
\begin{eqnarray}
\label{PrimalStateAs}
    1-\mathcal{W}_F(\mathcal{A}) = \max && \frac{1}{|X|} \sum_{i,x}\text{tr}[\sigma_{i|x}] \\
\text{s.t.: }&& p(i,x)\varrho_{i|x}\geq\sigma_{i|x}\ \forall i,x,\\\notag
&& \{\sigma_{i|x}\}_{i,x}\in C_F\notag,
\end{eqnarray}
where $|X|$ is the number of states for each $x$. This optimisation problem is an example of a cone program, the dual of which is given by~\cite{Gaertner2012, boyd2004convex}
\begin{eqnarray}
\label{DualStateAs}
    1-\mathcal{W}_F(\mathcal{A}) = \min_{Y\geq0}\, && \sum_{i,x}p(i,x)\tr{\varrho_{i|x}Y_{i|x}} \\
\text{s.t.: }&& \sum_{i,x}\tr{T_{i|x}Y_{i|x}}\geq1\ \forall \ \{T_{i|x}\}\in F\notag
\end{eqnarray}
where $Y=\bigoplus_{i,x}Y_{i|x}$ is a witness. The solution of the dual problem equals that of the primal given that the so-called Slater condition holds, which in this case can be verified by choosing $Y=\alpha\openone$ for large enough $\alpha>0$. This removes the redundant parts of the free cone as well.

To see the objective function of the dual problem as an instance of an input-output game, we define another witness as $\tilde Y_{i|x}:=Y_{i|x}/N$ where $N:=\|\sum_{i,x}Y_{i|x}\|$. For each $x$ we can add an extra term to $\{\tilde Y_{i|x}\}_i$, namely $\openone-\sum_{i,x}\tilde Y_{i|x}$, which ensures that we get a witness corresponding to a set of POVMs. Note that in the process we embed the state assemblage back into the larger space if needed and complete the witness into a POVM on the larger space by adding the missing parts to the last outcome. The new witness results in an objective function that is a scaled version of the success probability in a specific minimum error discrimination task. More precisely, $p_{succ}(\mathcal{A},\mathcal{M}_A):=\sum_{i,x}p(i,x)\tr{\varrho_{i|x}M_{i|x}}$. Clearly $p_{succ}$ is linear in the first argument and so from Eq.~\eqref{WeightDev} we get
\begin{align}
    p_{succ}(\mathcal{A},\mathcal{M}_A)\geq[1-\mathcal{W}_F(\mathcal{A})]\min_{\mathcal{A}_F\in F}p_{succ}(\mathcal{A}_F,\mathcal{M}_A),
\end{align} 
Noting that the scaling caused by adjusting the witness does not affect the following expression, putting the above inequality together with \eqref{DualStateAs} one arrives at
\begin{equation}
\label{RatioStateAs}
    \inf_{\mathcal M_A}\frac{p_{succ}(\mathcal{A},\mathcal{M}_A)}{\min_{\mathcal{A}_F\in F}p_{succ}(\mathcal{A}_F,\mathcal{M}_A)}=1-\mathcal{W}_F(\mathcal{A}),
\end{equation}
where we have used the standard convention that the optimisation is performed over those measurement assemblages $\mathcal M_A$ for which the l.h.s. is finite. We are ready to state our first observation.

\textit{Observation 1.} Let $F$ be a convex subset of state assemblages. For any state assemblage $\mathcal{A}\notin F$ there exists a set of measurements that anti-distinguishes the assemblage better than any assemblage in $F$. Moreover, the relative advantage is exactly quantified by the convex weight of $\mathcal{A}$ with respect to $F$.

As possible examples of the set $F$ we mention unsteerable assemblages and their generalisation to assemblages that can be prepared by states with an upper-bounded Schmidt number. In this case, the anti-distinguishing POVM can be alternatively interpreted as an instance of the task of subchannel exclusion supported by one-way local operations and classical communication (one-way LOCC), see \cite{Piani2015} for the details of such interpretation. In the case of state ensembles (i.e. assemblages with only one input $x=1$), the anti-distinguishing POVM relates to the task of subchannel exclusion on a single system, see \cite{Ryuji2019,uola2019a} for details. In the case of single states, one can relate the POVM to the task as a phase exclusion, see \cite{uola2019a}. In these cases, the possible examples include trivial ensembles, separable states, and states with a positive partial transpose.

We note that in the case where the free set $F$ consists of trivial ensembles, i.e. ensembles that code no information about which state was sent $\mathcal E:=\{\frac{1}{|X|}\varrho\}$, the weight corresponds to a measure of anti-distinguishability. In other words, the optimisation constraints are equivalent to the conditions $Y\geq0$ and $\sum_{i}Y_i\geq|X|\openone$. One can rewrite the second constraint as $\sum_{i}Y_i\geq\openone$ by multiplying the object function with $|X|$. Moreover, assuming that there is an optimal witness $\{\tilde Y_i\}$ that does not satisfy the equality $\sum_{i}\tilde Y_i=\openone$, one can filter the witness with the inverse square root of $\sum_i\tilde Y_i$, hence, ending up with the optimisation constraints that correspond to POVMs. Note that the weight with respect to a trivial ensemble gives a tight lower bound for the guessing probability, whereas the related generalised robustness measure gives a tight upper bound. Interestingly, if one wishes to approach either problem analytically, instead of searching for a POVM that optimally (anti-)distinguishes an ensemble, one can search for a single state for the trivial ensemble that optimises the corresponding convex distance, see Appendix for more details.

For a measurement assemblage $\mathcal{M}_A=\{M_{i|x}\}$ the primal problem reads
\begin{eqnarray}
\label{PrimalMeasAs}
   1-\mathcal{W}_F(\mathcal{M}_A) = \max && \frac{1}{|X|d} \sum_{i,x}\text{tr}[O_{i|x}] \\
\text{s.t.: }&& M_{i|x}\geq O_{i|x}\ \forall i,x,\\\notag
&& \{O_{i|x}\}_{i,x}\in C_F\notag,
\end{eqnarray}
where $d$ is the dimension of the Hilbert space. The dual of this reads
\begin{eqnarray}
\label{DualMeasAs}
    1-\mathcal{W}_F(\mathcal{M}_A) = \min_{Y\geq0}\, && \sum_{i,x}\tr{M_{i|x}Y_{i|x}} \\
\text{s.t.: }&& \sum_{i,x}\tr{T_{i|x}Y_{i|x}}\geq1\ \forall \ \{T_{i|x}\}\in F\notag
\end{eqnarray}
where $Y=\bigoplus_{i,x}Y_{i|x}$ is again a witness. With a similar argument as in the case of state assemblages, one checks that the Slater condition is valid. To get an expression similar to Eq.~\eqref{RatioStateAs}, we decompose the witness as $Y_{i|x}=\tilde N p(i,x)\varrho_{i|x}$, where $\varrho_{i|x}:=Y_{i|x}/\tr{Y_{i|x}}$, $p(i,x):=\tr{Y_{i|x}}/\tilde N$, and $\tilde N:=\sum_{i,x}\tr{Y_{i|x}}$. Noting again that scaling does not affect the desired expression, we write
\begin{equation}
\label{RatioMeasAs}
    \inf_{\mathcal A}\frac{p_{succ}(\mathcal{A},\mathcal{M}_A)}{\min_{\mathcal{O}_A\in F}p_{succ}(\mathcal{A},\mathcal{O}_A)}=1-\mathcal{W}_F(\mathcal{M}_A),
\end{equation}
where the optimisation is performed over those state assemblages $\mathcal{A}$ for which the l.h.s. is finite. We arrive at our second observation.

\textit{Observation 2.} Let $F$ be a convex subset of sets of POVMs. For any set of POVMs $\mathcal{M}_A\notin F$ there exists a state exclusion task where $\mathcal{M}_A$ outperforms any set of POVMs in $F$. Moreover, the relative advantage is exactly quantified by the convex weight of $\mathcal{A}$ with respect to $F$.

As examples of convex sets of measurements we mention joint measurability, informativity of a POVM (see Appendix for the explicit form of the weight), and simulability with projective (or any fixed subset of) POVMs. Note that in the case of sets of POVMs, the task includes the classical communication of the label $x$ from the preparing party to the measuring party. This allows the measuring party to choose the measurement setting after receiving the label. In the case of discrimination tasks, such scenario is referred to as state discrimination with pre-measurement information \cite{CaHeTo2019}.

In the case of quantum channels, i.e. completely positive trace-preserving maps, we start by writing the cone program in the Choi picture
\begin{eqnarray}
\label{WeightChanChoi}
1-\mathcal{W}_F(\Lambda) = \max\, && \tr{J_{\hat\Gamma}} \\
\text{s.t.: }&& J_\Lambda-J_{\hat\Gamma}\geq0,\quad J_{\hat\Gamma}\in C_{J_F}, \notag
\end{eqnarray}
where $J_\Lambda$ is the Choi state of $\Lambda$, and similarly for $\hat\Gamma$. 
The above optimisation problem is an instance of a cone program. Such a program comes with a dual formulation given by
\begin{eqnarray}
\label{WeightChanDual}
1-\mathcal{W}_F(\Lambda) = \min_{Y}\, && \tr{Y J_{\Lambda}} \\
\text{s.t.: }&& Y \geq 0, \quad \tr{YT}\geq 1\, \forall\ T\in J_{F}, \notag
\end{eqnarray}
where $Y$ is a dual variable constituting a witness for the set $J_F$. Once again, the Slater condition can be validated as in the case of state assemblages.

One can decompose the witness as $Y=d\sum_{ij}\omega_{ij}\ p(i)\varrho_i^T\otimes M_j$ for some state ensemble $\{p(i)\varrho_i\}$, POVM $\{M_j\}$, and set of real numbers $\{\omega_{ij}\}$. 
(The transpose is taken in the computational basis.) 
This decomposition shows that the weight $\mathcal{W}_F(\Lambda)$ is related to a payoff $P(\Lambda,\mathcal{G})$ of a specific input-output game:
\begin{align}
\label{Payoffwitness}
    \tr{Y J_{\Lambda}}&=\sum_{ij}d\ \omega_{ij}\  p(i)\tr{(\varrho_i^T\otimes M_j)J_\Lambda}\nonumber\\
    &=\sum_{ij}d\ \omega_{ij}\  p(i)\tr{\tr[\hil]{(\varrho_i^T\otimes I_\kil)J_\Lambda}M_j}\nonumber\\
    &=\sum_{ij}\omega_{ij}\  p(i) \tr{\Lambda(\varrho_i)M_j}\nonumber\\
    &=P(\Lambda,\mathcal{G}),
\end{align}
where in the penultimate inequality we have used the identity \eqref{eq:Choiinvert} given in the Appendix.

To get our result for channels, note that an optimal decomposition for $\Lambda$ from Eq.~(\ref{WeightDev}) with devices $D_F=\Gamma$ and $\tilde D = \tilde \Lambda$ gives a lower bound for the payoff of any canonical input-output game as
\begin{align}
\label{Payofflowerbound}
    P(\Lambda,\mathcal{G})&=[1-\mathcal{W}_F(\Lambda)]P(\Gamma,\mathcal{G}) + \mathcal{W}_F(\Lambda)P(\tilde\Lambda,\mathcal{G})\nonumber\\
    &\geq[1-\mathcal{W}_F(\Lambda)]P(\Gamma,\mathcal{G})\nonumber\\
    &\geq[1-\mathcal{W}_F(\Lambda)]\min_{\Gamma\in F}P(\Gamma,\mathcal{G}).
\end{align}
Note further that an input-output game given by an optimal witness $Y$ is up to scaling in the canonical form. This can be seen by putting Eq.~(\ref{WeightDev}) into the Choi picture and applying an optimal witness on both sides of the resulting equation. It follows that the payoff for the channel $\tilde\Lambda$ is zero. Putting this together with Eqs.~(\ref{WeightChanDual},\ref{Payoffwitness},\ref{Payofflowerbound}) and noting that the last equation is invariant under scaling of the games we get
\begin{align}
    \inf_{\mathcal{G}}\frac{P(\Lambda,\mathcal{G})}{\min_{\Gamma\in F}P(\Gamma,\mathcal{G})}=1-\mathcal{W}_F(\Lambda),
\end{align}
where the infimum is taken over all canonical input-output games. We are ready to state our third result.

\textit{Observation 3.} Let $F$ be a convex subset of channels. For any channel $\Lambda\notin F$ there exists an input-output game in which the channel results in a lower payoff than any channel in $F$. Moreover, the relative advantage is exactly quantified by the convex weight of $\Lambda$ with respect to $F$.

Note that this Observation can be directly generalised to the level of sets of channels and sets of quantum instruments
by considering the involved completely positive maps as a direct sum and having individual input-output games for each block. More precisely, for a set of instruments $\textbf{I}:=\{\mI_{i|x}\}_{i,x}$ one gets an extra coefficient $1/|X|$ and the Choi states become direct sums of the individual (subnormalised) ones, i.e. $\bigoplus_{i,x}J_{\mI_{i|x}}$ in Eq.~(\ref{WeightChanChoi}). The dual is simply a direct sum of the duals of the form Eq.~(\ref{WeightChanDual}) and the witnesses get the decomposition $Y_{i|x}=\sum_{a,b}p(a,i,x)\omega_{abix}\varrho_{a|i,x}\otimes M_{b|i,x}$. The payoff is then defined as the sum of all individual payoffs
\begin{align}
P(\textbf{I}, \mathcal{G}):=\sum_{a,b,i,x}p(a,i,x)\omega_{abix}\text{tr}[\mI_{a|x}(\varrho_{i|x,a}) M_{j|x,a}].
\end{align}
Using Eq.~(\ref{WeightChanDual}) it is straightforward to show that
\begin{align}
    \inf_{\mathcal{G}}\frac{P(\textbf{I}, \mathcal{G})}{\min_{\bm{\Gamma}\in F}P(\bm\Gamma,\mathcal{G})}=1-\mathcal{W}_F(\textbf{I}).
\end{align}

As examples of free sets $F$ we mention entanglement breaking channels, incompatibility breaking channels, compatible channels, compatible instruments, random unitaries, and finite rounds of LOCC protocols.

\textit{Conclusions.---} We showed that the convex weight, a natural measure for quantum resrouces, has all the desirable properties. Besides the basic requirements of faithfulness, convexity, and monotonicity, the convex weight also exactly captures the relative advantage of a quantum resource in an exclusion (or anti-distinguishability) task. 
%More importantly, we have shown that in contrast to commonly considered robustness-based quantifiers that highlight the outperformance of classical devices by quantum ones, our quantifier relates to the worst-case performance in terms of antidistinguishability.
This correspondance is fully general and can be applied in principle to any type of quantum resource. As examples we have discussed the cases of state assemblages, sets of POVMs and sets of transformations. 

Moreover, these ideas could be directly applied to experiments (similarly to those of Refs \cite{Sun18,Zheng2018}), as the exclusion task requires only control of the input state and the output measurements.
%In these scenarios the quantum properties have formerly gained interest in terms of the provided outperformance. For example, in the case of coherence and steering, such results have gained a considerable amount of attention from both the theoretical and experimental side.
%Our results show that one can equally well use antidistingushability for characterising and quantifying the quantumness of the involved devices in all of the mentioned scenarios.

\textit{Acknowledgments.---} The authors from the Turku Centre for Quantum Physics acknowledge financial support from the Academy of Finland via the Centre of Excellence program (Project No. 312058) as well as Project No. 287750. TB acknowledges financial support from the Turku Collegium for Science and Medicine.
TK acknowledges support by the DFG and the ERC (Consolidator Grant 683107/TempoQ). RU is grateful for the financial support from the Finnish Cultural Foundation. NB acknowledges support from the Swiss National Science Foundation (Starting grant DIAQ).

\textit{Note added.---} While finishing this manuscript, we became aware of the related work by A. Ducuara and P. Skrzypczyk \cite{1908.10347,upcoming}

\bibliography{bibliography.bib}

\appendix*
\section{Appendix}
{\it Convexity of the Weight.---} Consider two devices $D$ and $E$ and their mixture $pD+(1-p)E$, with $p\in[0,1]$. One can write
\begin{eqnarray}
    D&=&(1-\omega)D_F+\omega\Tilde{D}, \\
    E&=&(1-\nu)E_F+\nu \Tilde{E},
\end{eqnarray}
where this denotes the optimal solutions to the optimisation problem in Eq.~\eqref{WeightDev}, i.e., $\mathcal{W}(D)=\omega$ and $\mathcal{W}(E)=\nu$. We define new variables by
\begin{eqnarray}
\mu &=& p\omega + (1-p)\nu \\
M_F &=& \frac{p(1-\omega)D_F + (1-p)(1-\nu) E_F}{1-\mu}\\
\Tilde{M} &=& \frac{p\omega \Tilde{D} + (1-p)\nu \Tilde{E}}{\mu}
\end{eqnarray}
This allows us to write the mixture $pD+(1-p)E=(1-\mu)M_F+\mu \Tilde{M}$. The weight of the mixture $\mathcal{W}(pD+(1-p)E)$, is given by an optimal triplet $(\Tilde{\mu},N_F,\Tilde{N})$ for which
\begin{equation}
   pD+(1-p)E = (1-\Tilde{\mu})N_F+\Tilde{\mu}\Tilde{N}.
\end{equation}
From the optimality of $\Tilde{\mu}$ we can conclude that $\Tilde{\mu}\leq \mu$ and thus $\mathcal{W}(pD+(1-p)E)\leq p\mathcal{W}(D)+(1-p)\mathcal{W}(E)$.

\textit{Convex weight for transformations---}
Here we provide the technical details needed for the derivation of the convex weight in the case of transformations, with special attention being given to channels.

Consider a pair of Hilbert spaces $\hil, \kil$---with their respective spaces $\mathcal{L}(\hil),\mathcal{L}(\kil)$ of operators---and the set $T(\hil,\kil)$ of all transformations, i.e., completely positive trace-non-increasing maps, $\mI:\mathcal{L}(\hil)\rightarrow\mathcal{L}(\kil)$.
In the case that a given transformation is trace-preserving, we refer to it as a channel and denote it by $\Lambda$; the set $C(\hil,\kil)$ of all channels from $\mathcal{L}(\hil)$ to $\mathcal{L}(\kil)$ is a strict subset of $T(\hil,\kil)$.
It is important to note for our purposes that $T(\hil,\kil)$ is convex. 
For a given channel $\mI\in T(\hil,\kil)$ we define the \emph{Choi matrix} $J_{\mI}\in \mathcal{L}(\hil\otimes\kil)$ via $J_{\mI} = (id\otimes \mI)\proj{\psi^+}$, where $\ket{\psi^+}= 1/\sqrt{d}\ \sum_i \ket{i i} \in\hil\otimes\hil$ and $d=\dim(\hil)$. 
In the case that we are dealing with a channel $\Lambda$ the Choi matrix $J_\Lambda$ is both positive and of unit trace (for a transformation it is just positive), hence it is a state on $\hil\otimes\kil$ known as the \emph{Choi state} of $\Lambda$.
We may retrieve the original transformation from the Choi matrix via
\begin{equation}\label{eq:Choiinvert}
    \mI(\rho) = d \ \tr[\hil]{(\rho^T\otimes \id_\kil)J_{\mI}},
\end{equation}
where $\tr[\hil]{\cdot}$ denotes the partial trace over $\hil$ and the $T$ superscript denotes transposition with respect to the computational basis.
This corresponds to an inverse of the map $\mI \mapsto J_{\mI}$, and hence the map is an isomorphism.

We denote by $F\subset C(\hil,\kil)$ the set of \emph{free channels}, which are defined to be the channels satisfying $\mathcal{R}(\Lambda)=0$ for some resource measure $\mathcal{R}:C(\hil,\kil)\rightarrow \R$.
The free set varies depending on the intended resource---with examples being entanglement breaking, unitary and LOCC channels---though the most important point from our perspective is that such sets are themselves convex.
We denote by $C_{J_F} = \{t J_\Gamma | t\geq 0, \Gamma\in F \}$ the cone spanned by the image of the set $F$ under the Choi isomorphism.

For a given free set $F$, the \emph{convex weight} $\mathcal{W}_F$ of a channel is the largest degree to which that channel can be seen as an element of $F$. 
More explicitly, it is given by
\begin{equation}
\label{WeightChanPrimal}
\begin{split}
    \mathcal{W}_F(\Lambda) = &\min \lambda\\
    &\mathrm{s.t.:}\  \Lambda= (1-\lambda) \Gamma + \lambda \Tilde{\Lambda},\\
    & \Gamma\in F, \quad \Tilde{\Lambda}\in C(\hil,\kil). 
\end{split}
\end{equation}
If we apply the Choi isomoprhism, then Equation \eqref{WeightChanPrimal} can be rewritten in terms of the corresponding Choi states:
\begin{equation}
    \begin{split}
        \mathcal{W}_F(\Lambda) = &\min \lambda\\
    &\mathrm{s.t.:}\ J_\Lambda = (1-\lambda) J_\Gamma + \lambda J_{\Tilde{\Lambda}}, \\
    & J_\Gamma\in J_F, \quad \Tilde{\Lambda}\in C(\hil,\kil).
    \end{split}
\end{equation}
We wish to remove $J_{\Tilde{\Lambda}}$, and so we rearrange to $J_{\Tilde{\Lambda}} = \frac{1}{\lambda} (J_\Lambda - J_{\hat{\Gamma}})$, where $J_{\hat{\Gamma}}= (1-\lambda) J_\Gamma = J_{(1-\lambda) \Gamma}$.
The restrictions on the convex weight are now that $J_{\Tilde{\Lambda}}$ must be positive, corresponding to the equivalent condition of complete positivity of $\Tilde{\Lambda}$, and that $J_{\hat{\Gamma}}\in C_{J_F}$.
Noting that $1-\lambda$ is a nonnegative quantity, and that $1-\lambda = \tr{J_{\hat{\Lambda}}}$, we arrive at the desired form of the convex weight for the case of channels:
\begin{align*}
   1- \mathcal{W}_F(\Lambda) = &\max \tr{J_{\hat\Gamma}} \\
    &\mathrm{s.t.:}\  J_\Lambda-J_{\hat\Gamma}\geq0,\quad J_{\hat\Gamma}\in C_{J_F}.
\end{align*}

\textit{Finding the convex components of a POVM and bounding the convex weight.---} To give an analytical method for finding bounds on the convex weight, we present a characterisation of all POVMs together with the respective weight that can appear in a convex decomposition of a given POVM. The technique is based on the minimal Naimark dilation and although we present it only for the discrete case, we note that the continuous case can be treated in a similar manner by using the techniques in Ref.~\cite[Theorem~1]{Pellonp2012}.

We begin by recalling the a characterisation of non-normalised positive operator measures that are upper bounded by a POVM~\cite[Lemma~1]{Pellonp2014}. For this purpose, we fix a POVM $M$ with the input sigma-algebra $\Sigma$ and let $\big(\hil_{\oplus},J,P\big)$ be its minimal (diagonal) Naimark dilation. Especially, $J^*J=\id_\hil$. We let $D(\hil_\oplus)$ denote the algebra of (bounded) operators commuting with the projections $P_X$ for all $X\in\Sigma$. Let $P^J=JJ^*$ be the (minimal) Naimark projection from $\hil_\oplus$ onto the (closed) subspace $P^J\hil_\oplus$ of $\hil_\oplus$; we have $\hil\cong P^J\hil_\oplus$. 
\begin{lemma}
Let $N$ be a (possibly non-normalized) positive operator measure. Then $N_X\le M_X$ for all $X\in\Sigma$ if and only if there exists a (unique) $E\in D(\hil_\oplus)$, $0\le E\leq\openone_{\hil_\oplus}$, such that 
$N_X=J^*P_X EJ$ for all $X\in\Sigma$.
\end{lemma}

As a direct consequence we get can write for any POVM $M_1$ that is in a convex decomposition of $M$, i.e. $M_X=\lambda M_{X|1}+(1-\lambda)M_{X|2}$ for all $X\in\Sigma$ with $\lambda\in(0,1]$, the following result.
\begin{theorem}
There exists a unique $E_1\in D(\hil_\oplus)$ with $0\leq E_1\leq\lambda^{-1}\openone_{\hil_\oplus}$ and $J^* E_1 J=\openone_\hil$ such that $M_{X|1}=J^* P_XE_1J$ for all $X\in\Sigma$.
\end{theorem}

Motivated by the theorem we define a mapping $T:D_{P^J}^+(\hil_\oplus)\rightarrow \text{Comp}(M)$ by $T(E)_X=J^*P_X EJ$, where
\begin{align}
D_{P^J}^+(\hil_\oplus):=\{E\in D(\hil_\oplus)\,|\,E\geq 0,\;P^JE P^J=P^J\}    
\end{align}
and $\text{Comp}(M)$ consists of POVMs $M_1$ for which there exists another POVM $M_2$ and a weight $\lambda\in(0,1]$ such that $M_X=\lambda M_{X|1}+(1-\lambda)M_{X|2}$ for all $X\in\Sigma$; we say that $M_1$ is a component of $M$ with the weight $\lambda$.
Note that the condition $P^JE P^J=P^J$ is equivalent to $J^* E J=\openone_\hil$ and that for all $E\in D_{P^J}^+(\hil_\oplus)$ one has $\|E\|^{-1}\leq 1$. Hence, for any $E\in D_{P^J}^+(\hil_\oplus)$ one can take a $\lambda$ such that $E\leq\lambda^{-1}\openone_{\hil_\oplus}$ and write
\begin{align}
    M_X=\lambda T[E]_X + (1-\lambda) Q_X,
\end{align}
where $Q_X:=\frac{1}{1-\lambda}\big[M_X-\lambda T[E]_X\big]$ clearly normalised to identity and is seen to be positive by writing $M$ using the minimal dilation.

We are ready to state the following Corollary.
\begin{corollary}
Any number $\lambda\in(0,1]$ is a weight of a component $M_1$ if and only if $\lambda\leq\|E_1\|^{-1}$. Moreover, $M$ is extreme if and only if $D_{P^J}^+(\hil_\oplus)=\{\openone_{\hil_\oplus}\}$.
\end{corollary}

As an example of the above Corollary we take the case where the free set $F$ consists of trivial POVMs, i.e. POVMs of the form $O_i=p(i)\openone$ with $\{p(i)\}_i$ being a probability distribution. It is straight-forward to verify that a minimal dilation of a discrete POVM $M$ can be given through the isometry $J=\sum_{i,k}|e_{ik}\rangle\langle d_{ik}|$, where $\{|e_{ik}\rangle\}_{ik}$ is an orthonormal basis of the dilation space $\hil_\oplus=\bigoplus_i\hil_i$,  $\hil_i={\rm span}\{e_{ik}\}$,
 and $M_i=\sum_k|d_{ik}\rangle\langle d_{ik}|$ is the spectral decomposition of $M_i$ with orthogonal vectors $|d_{ik}\rangle\ne0$ (the eigenvalues are $\lambda_{ik}=\|d_{ik}\|^2$).

In order to bound the convex weight of $M$ with respect to the set $F$ we take a POVM $O\in F$ and write it using the above minimal dilation as
\begin{align}
    O_i=\sum_{kl}\langle e_{il} |E_i e_{ik}\rangle |d_{il}\rangle\langle d_{ik}|
\end{align}
where $E_i\in{\cal L}(\hil_i)$, $E_i\ge 0$.
Solving the matrix elements gives 
\begin{align}
    \langle e_{il} |E_i e_{ik}\rangle=\frac{p(i)}{\lambda_{ik}}\delta_{kl},
\end{align}
where $\{\lambda_{ik}\}_k$ are the eigenvalues of the POVM element $M_i$. Now the operator $E=\sum E_i$ has the norm $\|E\|=\sup_{ik}\{p(i)\lambda_{ik}^{-1}\}$. Hence, any point from the free set gives an upper bound on the convex weight. To see this, we note that $\sup_{ik}\{p(i)\lambda_{ik}^{-1}\}=\sup_i\{p(i)\lambda_{\min(M_i)}^{-1}\}$, where $\lambda_{\min(M_i)}$ refers to the smallest eigenvalue. The optimisation over the free set corresponds to optimising over the distributions $\{p(i)\}$. It is easy to check that for a distribution that maximises the weight of a trivial component, i.e. minimises $\sup_i\{p(i)\lambda_{\min(M_i)}^{-1}\}$, one needs to have $p(i)\lambda_{\min(M_i)}^{-1}=p(j)\lambda_{\min(M_j)}^{-1}$ for all $i,j$. Namely, were this not the case for an optimal distribution, one could define another distribution that replaces the $p(i)$ and $p(j)$ that give the highest and the second highest value of $p(i)\lambda_{\min(M_i)}^{-1}$ with $\tilde p(i)=\tilde p(j):=\frac{p(i)+p(j)}{2}$. Hence, the optimal distribution is $p(i)=\lambda_{\min(M_i)}/\sum_j \lambda_{\min(M_j)}$ the trivial weight is simply $1-W_F(M)=\sum_j \lambda_{\min(M_j)}$.

\textit{Finding the convex components of an instrument and bounding the convex weight.---} More generally, one can approach the convex weight of instruments and, hence, state ensembles in a similar manner~\cite[Example~2]{Pellonp2012}.
 A minimal Stinespring dilation of a (Heisenberg picture) instrument $\{\mathcal I_{i}^\dagger\}$ is given by an isometry $J$ and a normalized projection valued measure $\{P_i\}$ as
\begin{align}
    \mathcal I_i^\dagger(B)=J^\dagger (B\otimes P_i) J
\end{align}
where $B$ is a (bounded) operator of the output space. Recall that $\tr{\mathcal I_i^\dagger(B)\rho}=\tr{B\mathcal I_i(\rho)}$ where $\rho$ is an initial state.
It is straight-forward to check, using Lemma 1 of Ref.~\cite{Pellonp2014b}, that another instrument $\tilde{\mathcal I}_i^\dagger$ is a component of $\mathcal I_{i}^\dagger$ with weight $\lambda$ if and only if there exists a (unique) positive operator $E$ on the dilation space or ancilla $\hil_\oplus$, commuting with $\{P_i\}$, such that
\begin{align}
    \tilde{\mathcal I}_i^\dagger(B)=J^\dagger (B\otimes EP_i) J
\end{align}
with $J^\dagger (\openone\otimes E) J=\openone$ and $\lambda\leq\|E\|^{-1}$. Note that $E=\sum_i E_i$ where $E_i\ge0$ lives in the support space of $P_i$.
For example, consider a state ensemble $\{p(i)\varrho_i\}$. Now the input space is trivial (i.e.\ $\mathbb C$, $\cal L(\mathbb C)\cong\mathbb C$), $\mathcal I_i=p(i)\varrho_i$,  and ${\mathcal I}_i^\dagger(B)=\tr{p(i)\varrho_i B}=\langle\psi|(B\otimes P_i)\psi\rangle$ where $\psi$ is a purification of the total state $\sum_{i}p(i)\varrho_i=\tr[\hil_\oplus]{|\psi\rangle\langle\psi|}$ (since $J=|\psi\rangle\langle1|$) and $\{P_i\}$ constitutes a sharp resolution of the identity of the ancilla. 
Hence, the component $\tilde{\mathcal I}_i^\dagger(B)=\langle\psi^E|(B\otimes P_i)\psi^E\rangle$, where $\psi^E=(\id\otimes E^{1/2})\psi$ is a unit vector, so that $\tilde{\mathcal I}_i=\tilde p(i)\tilde\varrho_i=\tr[\hil_\oplus]{|\psi_i^E\rangle\langle\psi_i^E|}$ with
$\psi_i^E=(\id\otimes P_i)\psi^E=(\id\otimes E_i^{1/2})\psi$.
Especially, $\langle\psi_i^E|\psi_j^E\rangle=\tilde p(i)\delta_{ij}$ and $\tilde p(i)=\|(\id\otimes E_i^{1/2})\psi\|^2$ gives the necessary and sufficient conditions for the positive operators $E_i$:
$\sum_i\|(\id\otimes E_i^{1/2})\psi\|^2=\<\psi|(\id\otimes E)\psi\>=1$.
It is straight-forward to check that
\begin{align}
    E_i=\frac{\tilde p(i)}{p(i)}\varrho_i^{-1/2}\tilde\varrho_i\varrho_i^{-1/2}.
\end{align}
As a special free set, one can consider ensembles that carry no information about the sent state, i.e. $\tilde p(i)\tilde\varrho_i=\frac{1}{n}\tilde\varrho$. The corresponding convex weight corresponds to finding an optimal measurement for the task of exclusion. Hence, instead of searching for collections of positive operators (i.e. POVMs) that optimise the guessing probability, one can search for a single state that minimises the quantity $\max_i\|E_i\|$. We note that the task of minimum-error discrimination can be similarly be mapped into the search of a single state instead of a POVM, but in this case it is not clear how to find analytical bounds for the corresponding measure of generalised robustness.

\clearpage
\end{document}